\documentstyle[times,pramana,epsf,floats]{ias}
\def\beq{\begin{equation}}
\def\eeq{\end{equation}}
\def\bea{\begin{eqnarray}}
\def\eea{\end{eqnarray}}
\newcommand{\sla}{\!\!\!\!/ \,}

\begin{document}
\mark{{Some Applications....}{M. G. Mustafa}}
\title{ Some Applications of Thermal Field Theory to Quark-Gluon Plasma}

\author{Munshi G. Mustafa}
\address{Theory Group, Saha Institute of Nuclear Physics, 
1/AF Bidhan Nagar, Calcutta 700 064, India}
\keywords{Thermal Field Theory, Perturbative QCD, Hard Thermal Loop,  
Quark-Gluon Plasma}
\abstract{The lecture provides a brief introduction of thermal field 
theory within imaginary time formalism,
the Hard Thermal Loop perturbation theory and some of its application
to the physics of the quark-gluon plasma, possibly created in relativistic
heavy ion collisions.}

\maketitle
\section{Introduction}
The principle goal of the relativistic heavy-ion collision experiments is
the discovery of new state of matter, the so called quark gluon plasma
(QGP).
 The various measurements taken at CERN SPS 
 and at BNL RHIC do lead to strong $`$circumstantial evidence' for the formation of the
QGP~\cite{qm99,qm04}. Evidence
is circumstantial as any direct formation of the QGP cannot be identified.
Only by some noble indirect diagnostic probes like the suppression of the
$J/\Psi$ particle, the jet quenching,
the enhanced production of strange particles, specially strange antibaryons,
excess production of photons and dileptons, 
etc. the discovery can be achieved.


In order to understand the properties of a QGP and to make unambiguous 
predictions about signature of QGP formation, one needs a profound description 
of QGP. For this purpose we have to use QCD at finite temperature and
chemical potential. There are two different approaches:
1) Lattice QCD is a non-perturbative method for solving the QCD equations
numerically on a 4-dimensional space-time lattice~\cite{karsch0}.
In this way all temperatures from below to above the phase transition are
accessible. 
 2) Perturbative QCD at finite temperature~\cite{kapus,lebel} is based on 
the fact that the
temperature dependent running coupling constant is small at high
temperatures due to asymptotic freedom, $T\rightarrow \infty
\Rightarrow \alpha _s(T)=g^2/4\pi \rightarrow 0$. At a typical temperature
of $T=250$ MeV we expect $\alpha _s= 0.3$ -- 0.5. This suggests that
perturbation theory could work at least qualitatively. 
This corresponds to an
expansion in $\alpha_s$, which can be performed conveniently by using
Feynman diagrams for partonic scatterings.
\vspace{-0.4cm}
\section{Thermal Field Theory}
Thermal field theory is a combination of three basic branches of modern
physics, {\it viz.}, quantum mechanics, theory of relativity and statistical
physics. Our aim is to derive Feynman diagrams and rules at 
$T>0 \, (\mu\neq 0)$. 
We start by repeating some of the basic facts in equilibrium  statistical
mechanics. The statistical behaviour of a quantum system, in thermal
equilibrium, is studied through an appropriate ensemble and is defined as
\begin{equation}
\rho(\beta)= e^{-\beta {\cal H}} \, \, , \label{eq1}
\end{equation}
where $\beta=1/T$  (the
Boltzmann constant $k_B\equiv1$) and $\cal H$ is appropriate Hamiltonian
for a given choice of ensemble. 
Given the density matrix, the finite temperature behaviour of any theory 
is specified by the partition function
\begin{equation}
{\cal Z}(\beta)={\rm{Tr}} \rho(\beta) ={\rm{Tr}}e^{-\beta {\cal H}} \, \,
.\label{eq2}
\end{equation}

The thermal expectation value of any physical observable can be written as
\begin{equation}
\langle {\cal A}\rangle_\beta = {\cal Z}^{-1}(\beta) {\rm{Tr}}\left [ \rho(\beta)
{\cal A}\right ]={\cal Z}^{-1}(\beta) {\rm{Tr}}\left [
e^{-\beta {\cal H}} {\cal A} \right ] \, \, , \label{eq3}
\end{equation}
and that of correlation function of any two observables is given as
\begin{equation}
\langle {\cal A}{\cal B}\rangle_\beta = {\cal Z}^{-1}(\beta) {\rm{Tr}}
\left [ \rho(\beta) {\cal A}{\cal B}\right ]={\cal Z}^{-1}(\beta)
 {\rm{Tr}}\left [
e^{-\beta {\cal H}} {\cal A} {\cal B}\right ] \, \, , \label{eq4}
\end{equation}
For a given Schr\"odinger operator, ${\cal A}$, the Heisenberg operator,
${\cal A}_H(t)$ can be written as
\begin{equation}
{\cal A}_H(t) = e^{i{\cal H}t} \, {\cal A} \,  e^{-i{\cal H}t} \, \, .
\label{eq5}
\end{equation}
The thermal correlation function of two operators 
can also be written as
\begin{eqnarray}
\langle {\cal A}_H(t) {\cal B}_H(t') \rangle_\beta &=& {\cal Z}^{-1}(\beta) 
{\rm{Tr}}\left [ e^{-\beta {\cal H}} {\cal A}_H(t) 
{\cal B}_H(t') \right ] = 
\langle {\cal B}_H(t') {\cal A}_H(t+i\beta) \rangle_\beta
\, \, , \label{eq6}
\end{eqnarray}
which holds irrespective of Grassmann parities of the operators.
Eq.(\ref{eq6})
is known as the Kubo-Martin-Schwinger (KMS) relations and will
lead to (anti)periodicity in various 2-points functions at finite
temperature. 

We would like to note that in (\ref{eq2}) $``${\bf{Tr}}" indicates the
sum over expectation values in all possible states in the Hilbert space and 
there is  an infinite number of such basis in quantum field 
theory~\cite{Das,markus}. The 
partition function of a statistical system cannot be computed exactly even if
one makes a perturbative expansion into a power series in coupling constant, 
$g$. The Matsubara formalism, known as also imaginary time formalism, 
provides a diagrammatic way of computing the 
partition function and other relevant physical quantities perturbatively. 

\subsection{Matsubara formalism}

The Hamiltonian of a system can be decomposed as
\begin{equation}
 {\cal H} = { \cal H}_0 + {\cal H}' \, \, , \label{eq7}
\end{equation}
 where ${\cal H}_0$ and ${\cal H}'$ are the free and interaction parts, 
respectively. The density matrix in (\ref{eq1}) becomes
\begin{equation}
\rho (\beta) 
= \rho_0 (\beta) {\cal S}(\beta) \, \, \, \, 
{\rm{with}} \, \,  \rho_0 (\beta) \equiv e^{-\beta {\cal H}_0} \, \, \,
; \, \, \, 
{\cal S}(\beta) = e^{\beta {\cal H}_0} \ e^{-\beta {\cal H}}. \label{eq8}
\end{equation}
The density matrix can have evolution equation with $0\leq\tau\leq\beta$:
\begin{eqnarray}
 \frac{\partial \rho_0(\tau)}{\partial \tau} &=& -{\cal H}_0 \rho_0(\tau) \, 
\, \, {\rm{and}} \, \, \, 
\frac{\partial \rho(\tau)}{\partial \tau} = -{\cal H} \rho (\tau)
= - \left ({ \cal H}_0 + {\cal H}'\right ) \rho(\tau) \, . \label{eq9}
\end{eqnarray}
Now ${\cal S}(\tau)$ satisfies the evolution equation, following 
(\ref{eq8}) and (\ref{eq9}),  as
\begin{eqnarray}
 \frac{\partial {\cal S} (\tau)}{\partial \tau} &=&
\frac{\partial \rho_0^{-1}(\tau)}{\partial \tau} \rho(\tau)
+ \rho_0^{-1}\frac{\partial \rho(\tau)}{\partial \tau} 
= -{\cal H}'_I(\tau) {\cal S}(\tau) \, \, , \label{eq10}
\end{eqnarray}
with ${\cal H}'_I(\tau) = \rho_0^{-1}(\tau) {\cal H}' \rho_0(\tau)
= e^{\tau {\cal H}_0} {\cal H}' e^{-\tau {\cal H}_0} $ in a modified
interaction picture which is similar to {\it zero temperature field theory}. 
Note that for a real $\tau$ such
transformation may not be necessarily {\it unitary} but for 
{\it imaginary} values
of $\tau =it $ it will be so. This makes Matsubara formalism a {\it imaginary 
time} formalism. It is also to be noted that under such rotation to imaginary 
time the field remains {\it hermitian} with the appropriate definition of 
hermiticity for complex coordinates~\cite{Das}, 
{\it viz.}, $\phi^\dagger(z)=\phi(z^*)$. 
Integrating (\ref{eq10}) one can obtain ~\cite{Das}
\begin{equation}
 {\cal S}(\beta) = T_\tau \left ( e^{-\int_0^\beta {\rm d}\tau {\cal H}_I'
(\tau)} \right ) \, \, , \label{eq11} 
\end{equation}
 where $T_\tau $ is the ordering of $\tau$ variable. The following points 
to be noted:
1) Eq.(\ref{eq11}) resembles the $\cal S$-matrix in zero temperature 
field theory with the exception that the time integration becomes now finite.
2) Now one can expand the exponential in (\ref{eq11}) into a power series
of the coupling constant and each term would then correspond to a Feynman 
diagram. 
3) In this formalism Wick's theorem can easily be generalized to finite 
temperature. 

\subsection{Matsubara frequency}
The 2-point Greens functions can be defined in Heisenberg picture as
\begin{eqnarray}
{\cal G}_\beta(\tau,\tau')&=&\left \langle T_\tau\left ( \phi_H(\tau)\phi_H^\dagger
(\tau')\right ) \right \rangle_\beta ={\cal Z}^{-1}(\beta) {\rm{Tr}} e^{-\beta{\cal H}}
T_\tau\left ( \phi_H(\tau)\phi_H^\dagger (\tau')\right ) \, \, , \label{eq12}
\end{eqnarray}
with $\phi$ is a complex field and 
the $\tau$ ordering is defined as
\begin{eqnarray}
T_\tau\left ( \phi_H(\tau)\phi_H^\dagger (\tau')\right )&=&
\theta(\tau-\tau')\phi_H(\tau)\phi_H^\dagger (\tau')
\mp \theta(\tau'-\tau)\phi_H^\dagger(\tau')\phi_H (\tau) \, , \label{eq13}
\end{eqnarray} 
where negative sign in the second term is for fermionic field as the $\tau$
ordering is sensitive to the Grassmann parity of the fields. The spatial
dependence and spinorial indices are omitted for the time being as they are 
similar to the zero temperature field theory. 


Exploiting the cyclicity properties of the trace and (\ref{eq13}), 
one can show that for $\tau > 0$ (\ref{eq12}) satisfies (anti)periodic 
conditions
\begin{equation}
{\cal G}(0,\tau)=\mp\left \langle \phi_H^\dagger(\tau)\phi_H(0)
\right\rangle_\beta = \mp {\cal G}_\beta(\beta,\tau) \, \, , \label{eq15}
\end{equation}
for (fermion)boson which restrict the time in finite interval $[0,\beta]$.
The (\ref{eq15}) is equivalent to KMS relation 
defined in (\ref{eq6}) with a suitable rotation to imaginary times.

The Fourier transform of 2-point functions would involve the discrete
frequencies as those are defined in finite time interval and in general
can be written as
\begin{eqnarray}
 {\cal G}_\beta(\tau) &=& \frac{1}{\beta} \sum_n e^{-i\omega_n \tau}
{\cal G}_\beta(\omega_n) \, \, ,  \label{eq16} \\
{\cal G}_\beta(\omega_n) &=& \frac{1}{2} \int_{-\beta}^\beta  {\rm d}\tau
e^{i\omega_n \tau} {\cal G}_\beta(\tau) \, \, , \label{eq17}
\end{eqnarray}
where $\omega_n = n\pi/\beta$ with $n=0,\ \pm 1, \ \pm 2, \cdots $. 
The (anti)periodicity conditions in (\ref{eq15}) restricts 
(\ref{eq17}) as 
\begin{eqnarray}
{\cal G}_\beta(\omega_n)
&=& \frac{1}{2} \int_{-\beta}^0 {\rm d}\tau
e^{i\omega_n \tau} {\cal G}_\beta(\tau)
+ \frac{1}{2} \int_{0}^\beta {\rm d}\tau e^{i\omega_n \tau}
{\cal G}_\beta(\tau) \nonumber \\
&=& \frac{1}{2} \left (1 \mp (-1)^n \right ) \int_{0}^\beta {\rm d}\tau
e^{i\omega_n \tau} {\cal G}_\beta(\tau) \, \, , \label{eq18}
\end{eqnarray}
yielding discrete frequencies
$ \omega_n =
 \left \{ \begin{array}{ll} {2n\pi T} &\mbox{{for bosons}} \\
                           {(2n+1)\pi T} &\mbox{{for fermions}}
\end{array} \right.$
, also known as {\it Matsubara frequencies}.

{\it Two important consequences of imaginary time formalism}: 
1) The time $\tau$
is restricted to the finite interval $[0,\beta]$ due to 
(anti)periodic conditions
for (fermion)boson. 2) The finite time interval changes the integral over
the time component of momentum at zero temperature to a Fourier series
over  Matsubara frequencies.

\subsection{Propagator for any theory}
Since the spacial dependencies of 2-point functions are continuous like 
zero temperature field theory, therefore, putting all the coordinates 
together one can write
\begin{eqnarray}
{\cal G}_\beta({\vec x},\tau) &=& \frac{1}{\beta} \sum_n \int \frac{{\rm d}^3k}
{(2\pi)^3}e^{-i(\omega_n \tau-{\vec k \cdot \vec x)}} 
{\cal G}_\beta({\vec k},\omega_n)\, \, , \label{eq20a}\\
{\cal G}_\beta({\vec k},\omega_n) &=&  \int_{0}^\beta  {\rm d}\tau
\int {{\rm d}^3x}
e^{i(\omega_n \tau-{\vec k \cdot \vec x})} {\cal G}_\beta({\vec x},\tau) \, .
\label{eq20b}
\end{eqnarray}

The zero temperature Greens function satisfies
the bosonic Klein Gordon equation as 
\begin{equation}
\left ( \partial_\mu\partial^\mu+m^2)\right ){\cal G}(x)=\delta^4(x) \, ,
\label{eq21}
\end{equation}
with the choice of metric in Minkowski space is $(+,-,-,-)$.
While rotating to the imaginary time ($t\rightarrow -i\tau \, \, 
\Rightarrow \, \, {\cal G}\rightarrow -{\cal G}_\beta$) the finite
temperature Greens function satisfies the equation
\begin{equation}
\left ( \frac{\partial^2}{\partial \tau^2} + \nabla^2 - m^2 \right )
{\cal G}_\beta ({\vec x},\tau)= -\delta^3({\vec x}) \delta(\tau) \, \, .
\label{eq22}
\end{equation}
On substituting 
(\ref{eq20a}) in (\ref{eq22}), the momentum space propagator is obtained as
\begin{equation}
{\cal G}_\beta ({\vec k},\omega_n) = \frac{1}{\omega_n^2+{\vec k}^2+m^2} \,
\, . \label{eq23}
\end{equation}

Now substituting (\ref{eq23}) in (\ref{eq20a}) and performing
the frequency sum~\cite{lebel}, one can obtain
\begin{eqnarray}
{\cal G}_\beta({\vec x},\tau)&=& \
\int \frac{{\rm d}^3k}{(2\pi)^3} \frac{e^{\vec k \cdot \vec x}}{2\omega_k}
\left[\left (1+n_B(\omega_k)\right)e^{-\omega_k \tau}+ n_B(\omega_k)
e^{\omega_k \tau} \right ] \, \, , \label{eq25}
\end{eqnarray}
where $\omega_k=\sqrt{k^2+m^2}$ and $n_B(\omega_k)=(e^{\beta\omega_k}-1)^{-1}$
is distribution function for boson. Now putting $\tau=it$ and $n_B(\omega_k)=0$,
one can recover the zero temperature 2-point Greens function for this case. 
Similarly, one can also obtain 2-point Greens function for fermion from Dirac
equation.

\subsection{Feynman rules for finite temperature field theory}
\begin{enumerate}
\item Replace $T=0$ propagator by $T\neq 0$ propagator as obtained which
carry $T$ dependence through Matsubara frequency.
\item The vertex is same as $T=0$ case.
\item Replace loop integrals by $\int d^4K/(2\pi)^4 \rightarrow 
T\sum_n \int d^3k/(2\pi)^3\, \,$ .
\item Symmetry factors are same as $T=0$ case.
\end{enumerate}

\section{Hard Thermal Loop (HTL) Approximation}
Restricting only to bare propagators and vertices perturbative 
QCD can lead to serious problems, {\it i.e.}, infrared divergent and gauge 
dependent results 
for physical quantities, {\it e.g.}, in the case of
the damping rate of a long wave, 
collective gluon mode in the QGP. The sign and magnitude of the gluon
damping rate was found to be strongly gauge dependent~\cite{Kaj85,Lop85}.
The reason for such behaviour is the fact that bare perturbative QCD at 
finite temperature is incomplete, {\it i.e.}, higher order diagrams missing
in bare perturbation theory can contribute to lower order in coupling constant.
In order to overcome these problems, the HTL resummation technique, 
{\it an improved 
perturbation theory}, has been suggested by Braaten and 
Pisarski~\cite{Bra90a} and also by Frenkel and 
Taylor~\cite{Fre90}, in which those diagrams can be taken
into account by resummation. 
The starting point is
the separation of scales in the weak coupling limit, $g<<1$, since there are
two momentum scales in a plasma
of massless particles: {\bf (i)} {\it hard}, where the 
momentum $\sim$ temperature, $T$ and {\bf (ii)} {\it soft}, where the momentum 
$\sim $ thermal 
mass ($\sim \ gT$, $g<<1$). 

\vspace{-0.5cm}
\subsection{HTL perturbation theory (HTLpt)}
The HTL resummation technique allows a systematic gauge invariant
treatment of gauge theories at finite temperature and chemical potential
taking into account medium effects such as Debye screening, effective quark
masses, and Landau damping \cite{Bra90a}.
The generating functional which generates the HTL Green functions between
a quark pair and any number of gauge bosons can be written
\cite{Bra90a,braaten0} as
\begin{equation}
\delta{\cal L} = m_q^2 {\bar \psi} \left \langle
\frac{K \! \! \! \! /}{K\cdot D} \right \rangle \psi \ \ , \label{hg}
\end{equation}
where $K^\mu$ is a light like four-vector, $D_\mu$ is the covariant
derivative, $m_q$ is thermal quark mass, and $\langle \ \ \rangle$ is the
average
over all possible directions over loop momenta. This functional
is gauge symmetric and nonlocal and leads to the following
Dirac equation
\begin{equation}
D \! \! \! \! / \psi =  \Sigma \psi + \Gamma_\mu A^\mu \psi + A^\mu
\Gamma_{\mu\nu}A^\nu \psi
 +\cdots  \ \ \ , \label{hd}
\end{equation}
where we have suppressed the color index.
In the HTL approximation the 2-point function, $\Sigma\sim gT$,
(quark self-energy) is of
the same order as the tree level one, $S^{-1}_0(K) \sim K \! \! \! \! /
\sim gT$ (in the weak coupling limit $g\ll 1$),
if the external momenta are soft, {\it i.e.} of the order $gT$.
The 3-point function,
$i.e.,$ the effective quark-gauge boson vertex, is given by
$g\Gamma_\mu=g(\gamma_\mu+\delta \Gamma_\mu)$, where $\delta \Gamma_\mu$ is the
HTL correction.
The 4-point function, $g^2\Gamma_{\mu\nu}$,
does not exist at the tree level and only appears
within the HTL approximation \cite{Bra90a}.
These $N$-point functions, which are complicated functions of momenta and
energies, are interrelated by Ward identities. 
The HTLpt~\cite{andersen}
is an extension of the screened perturbation theory~\cite{karsch}
in such a way that
it amounts to a reorganization of the usual perturbation theory by
adding and subtracting the nonlocal HTL term (\ref{hg}) in the
action~\cite{Bra90a,Fre90}. The added
term together with the original QCD action is treated nonperturbatively
as zeroth order whereas the subtracted one as perturbation.
Equivalently, for calculating physical quantities this amounts to replace
the bare propagators and vertices by resummed HTL Green
functions~\cite{Bra90a}.
However, HTLpt suffers from the fact
 that one uses HTL resummed Green functions
also for hard momenta of the order $T$.
A systematic description of physical quantities requires an explicit
separation of hard ($\sim T$) and
soft ($\sim gT$) scales~\cite{Bra90a}. 
Following this approach, a number of relevant physical
quantities, such as signatures
of the QGP, has consistently been 
calculated~\cite{markus,andersen0,andersen1,andersen1a,andersen2,Bra90}.

\subsection{Quark self energy in HTL-approximation}
\vspace{-0.3in}
\begin{figure}[htbp]
\epsfxsize=8cm
\hspace*{-3cm}\centerline{\epsfbox{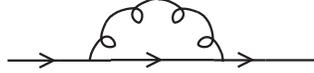}}
\caption{Quark self energy having quark momentum $K$.}
\label{fig:qself}
\end{figure}

The most general ansatz for the fermionic self energy 
in rest frame of the plasma is given by~\cite{Wel82a}  

\beq
\Sigma(K)=-a(k_0,k)K\sla -b(k_0,k)\gamma_0 \ , \label{eq26}
\eeq
where $K=(k_0,{\bf k})$,  $k=|{\bf k}|$ and the quark mass is neglected
assuming that the temperature is much larger than the quark mass, which holds 
at least for $u$ and $d$ quarks.  The scalar quantities $a$ and $b$ are
given by the traces over self energies within HTL approximations
 in figure~\ref{fig:qself} as, 
respectively,
\vspace{-0.0in}
\bea
a(k_0,k) & = & \frac{1}{4k^2}\> \left [{\rm {tr}}\, (K\sla \Sigma ) - k_0\>
{\rm{tr}}\, (\gamma _0
\Sigma )\right ]
=\frac{m_q^2}{k^2}\> \left (1-\frac{k_0}{2k}\> \ln
\frac{k_0+k}{k_0-k}\right )\ \ , \label{eq27} \\
b(k_0,k) & = & \frac{1}{4k^2}\> \left [K^2\> {\rm{tr}}\, (\gamma _0 \Sigma )
 - k_0\> {\rm{tr}}\,
(K\sla \Sigma )\right ]=
\frac{m_q^2}{k^2}\> \left (-k_0+\frac{k_0^2-k^2}{2k}\> \ln
\frac{k_0+k}{k_0-k}\right ) ,\label{eq28}
\eea
where $m_q^2={g^2T^2}/{6}$, is the effective quark mass. The quark self 
energy in (\ref{eq26}) has an imaginary part below the light cone,
$k_0^2-k^2<0$, representing {\it Landau damping (LD)}
 for space like quark momenta.
Furthermore, the general ansatz in (\ref{eq26}) is also chirally 
invariant in spite of the appearance of an effective quark mass.

\subsection{Effective propagators and vertices in HTL-approximation}
\vspace{-0.5cm}
\begin{figure}[htbp]
\epsfxsize=8cm
\centerline{\epsfbox{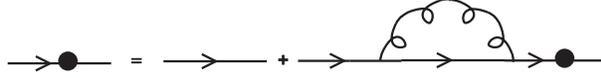}}
\caption{Effective quark propagator in HTL-approximation}
\label{fig:qprop}
\end{figure}
Resumming the quark self-energy by using
the Dyson-Schwinger equation, the effective quark propagator
in figure~\ref{fig:qprop} can be written as 
\beq
S(K)=[K\sla -\Sigma(K)]^{-1} 
= \frac{\gamma_0-\hat {\bf k}\cdot {\vec \gamma}}{2D_+(K)}+
\frac{\gamma_0+\hat {\bf k}\cdot {\vec \gamma}}{2D_-(K)} \ \ ,  \label{eq29}
\eeq
for massless quarks with a decomposition into helicity eigenstates, 
where
\beq
D_\pm(k_0,k)=(1+a(k_0,k))\> (-k_0\pm k)+b(k_0,k) \ \ \ . 
\label{eq31}
\eeq
\vspace{-0.3in}
\begin{figure}[htbp]
\epsfxsize=5cm
\centerline{\epsfbox{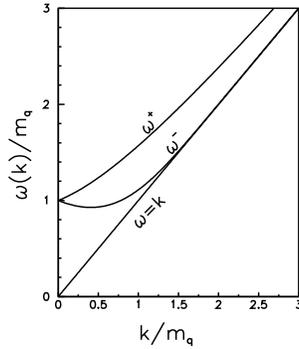}}
\vspace{-0.4in}
\caption{Quark dispersion relation in HTL-approximation along with the free 
($\omega=k$) one.}
\label{fig:hdisp}
\end{figure}

The relevant QED like HTL-vertex is related to this propagator
through the Ward identity~\cite{Bra90,Fre90} 
\beq
K_\mu \Gamma^\mu(K_1,K_2;K)=S^{-1}(K_1)-S^{-1}(K_2) \ \ . \label{eq32}
\eeq

Now, the zeros of $D_\pm(K)$ give the in-medium propagation or
{\it quasiparticle (QP)} dispersion relation. 
As shown in figure~\ref{fig:hdisp} the upper curve 
$\omega_{+}(k)$ corresponds to the solution of $D_{+}(K)=0$, whereas 
the lower curve $\omega_{-}(k)$ represents the solution of $D_{-}(K)=0$. 
Both branches start from a common effective
mass, $m_q$, obtained in the $k\rightarrow 0$ limit~\cite{Bra90}.
The $\omega_{+}(k)$ branch describes the propagation of an ordinary quark
with thermal mass, and
the ratio of its chirality to helicity is $+1$. On the other hand,
the $\omega_{-}(k)$ branch corresponds to the propagation of a quark mode
with a negative chirality to helicity ratio. This branch represents the
plasmino mode which is absent in the vacuum but appears as consequence of 
the medium due to the broken Lorentz invariance, and has a shallow 
minimum.  This corresponds to a purely collective long 
wave-length mode, whose spectral strength decreases exponentially at high 
momenta.  
For high momenta, however, both 
branches approach the free dispersion relation. 

The spectral representation~\cite{markus,Bra90} of 
HTL propagator in (\ref{eq29}) can now be written as
\begin{eqnarray}
 \rho_\pm &=&\frac{1}{\pi} {\rm {Im}} f_\pm(x+i\epsilon)= {\mathrm {Pole}}
(k_0^2>k^2)+{\mathrm{Cut}} (k_0^2<k^2) \, \, . \label{eq32a}
\end{eqnarray}

\section{Thermal Hadronic Correlation Function}
While lattice calculations of hadron properties in vacuum have reached
quite satisfactory precision, little is known from such first principle
calculations about basic hadronic properties in a thermal medium.
We want to analyse~\cite{Kar01}
its behaviour in the high temperature limit through the thermal correlation
functions in terms of quark fields.

\subsection{Definition}

Meson correlators are constructed from meson currents
$J_M (\tau,\vec{x}) =\bar{q}(\tau, \vec{x})\Gamma_M q(\tau, \vec{x})$,
where $\Gamma_M = 1$, $\gamma_5$, $\gamma_\mu$, $\gamma_\mu \gamma_5$ 
for scalar, pseudo-scalar, vector and pseudo-vector channels, respectively.
The thermal two-point functions in coordinate space, $G_M(\tau,\vec{x})$,
are defined as
\begin{eqnarray}
G_M(\tau,\vec{x}) &=&
\langle J_M (\tau, \vec{x}) J_M^{\dagger} (0, \vec{0}) \rangle
= T \sum_{n=-\infty}^{\infty} \int
{{\rm d}^3p \over (2 \pi)^3} \;{\rm e}^{-i(\omega_n \tau- \vec{p} \vec{x})}\;
\chi_M(\omega_n,\vec{p})~~,
\label{eq33}
\end{eqnarray}
where $\tau \in [0,1/T]$, and the Fourier transformed correlation function
$\chi_M(\omega_n,\vec{p})$ is given at the discrete Matsubara modes,
$\omega_n = 2n \pi T$. The imaginary part of the momentum space correlator
gives the spectral function $\sigma_M(\omega,\vec{p})$,
\begin{equation}
\chi_M(\omega_n,\vec{p}) = -\int_{-\infty}^{\infty} {\rm d}
\omega {\sigma_M(\omega,\vec{p}) \over i\omega_n - \omega +i\epsilon}
\Rightarrow 
\sigma_M(\omega,\vec{p}) = {1\over \pi} {\rm Im}\;  \chi_M(\omega,\vec{p}).
\label{eq34}
\end{equation}
Using (\ref{eq33}) and (\ref{eq34}) we obtain
the spectral representation of the thermal correlation function in
coordinate space at fixed momentum ($\beta =1/T$),
\begin{equation}
G_M(\tau,\vec{p}) =  \int_{0}^{\infty} {\rm d} \omega\;
\sigma_M (\omega,\vec{p})\;
{{\rm cosh}(\omega (\tau - \beta/2)) \over {\rm sinh} (\omega \beta/2)}~~.
\label{eq35}
\end{equation}

\subsection{Hadronic spectral and correlation function in HTL-approximation}

\begin{figure}[htbp]
\epsfxsize=4cm
\centerline{\epsfbox{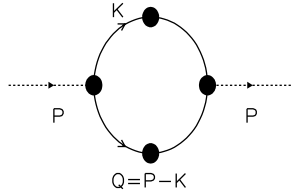}}
\caption{Self-energy diagram involving quarks loop in HTL-approximation.}
\label{fig:hself}
\end{figure}

The hadronic spectral functions, $\sigma_M(\omega, p)$ in (\ref{eq34}),
of the temporal correlators are proportional to the imaginary part of
the quark loop diagram.
The meson spectral function~\cite{Kar01}, constructed from two quark 
propagators, will have pole-pole, pole-cut and
cut-cut contributions,
$\sigma_M(\omega)=\sigma^{\rm{pp}}(\omega)+\sigma^{\rm{pc}}(\omega)
+\sigma^{\rm{cc}}(\omega)$.
Free meson spectral functions are also
obtained using bare propagators and vertices
in figure~\ref{fig:hself} as~\cite{Kar01,Flo94}
$\sigma ^{{\rm free}}_{M}\left( \omega \right)
=\frac{N_{C}}{4\pi ^{2}} \omega ^{2}\tanh \left( \frac{\omega }{4T}\right)
a_{M}$
where $a_M= (+)-1$ for (pseudo)scalar and $(-)+2$ for (pseudo)vector.

\begin{figure}[htbp]
\centerline{{\epsfxsize=6cm {\epsfbox{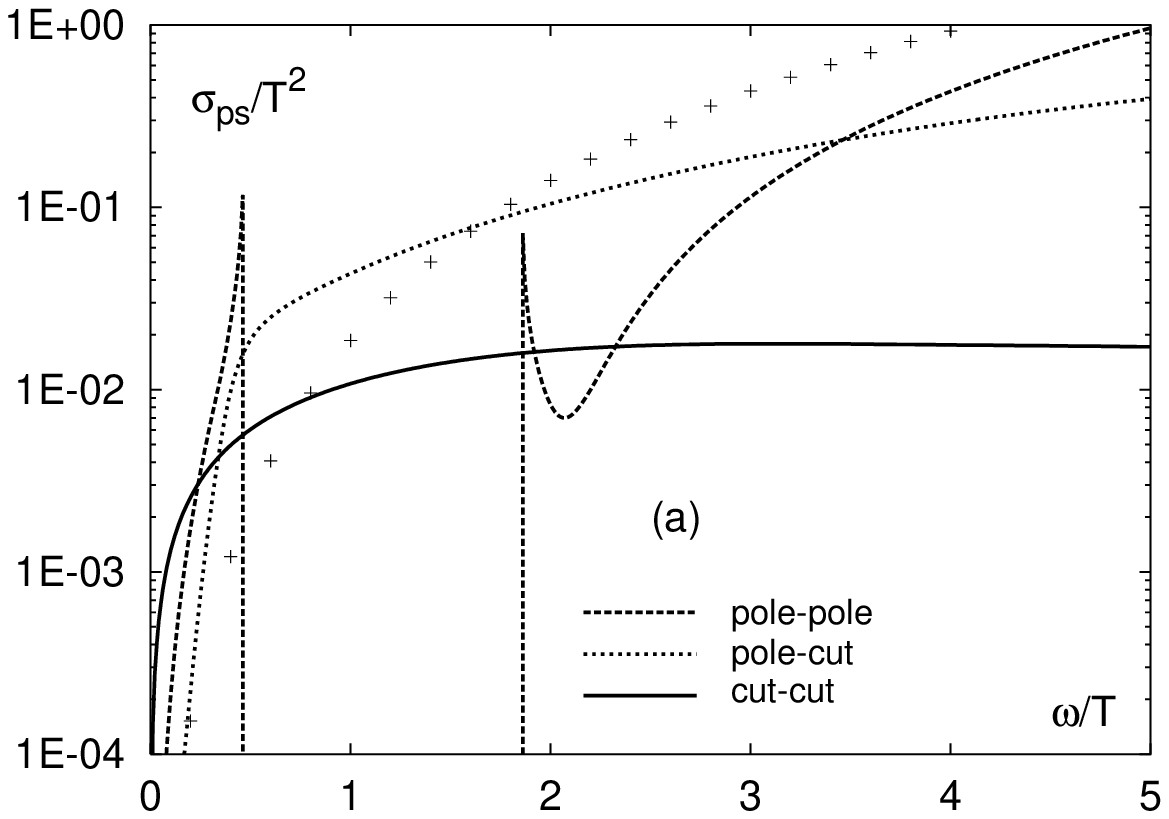}}}
{\epsfxsize=6cm {\epsfbox{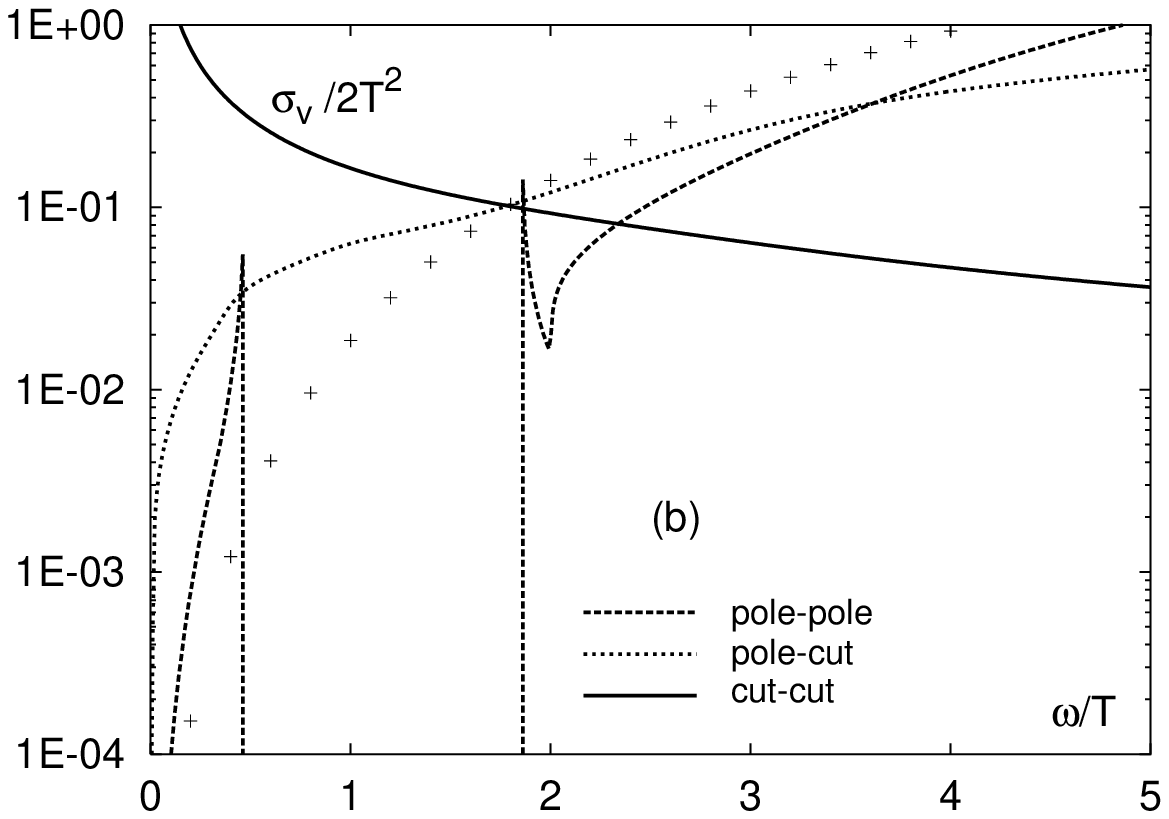}}}}
\caption{
(a) Pseudoscalar and (b) vector meson spectral functions in HTL approximation
for $m_q/T=1$. The free one is represented by crosses.}
\label{fig:hspec}
\end{figure}
In figure~\ref{fig:hspec} we show the different contributions for 
pseudoscalar and vector mesons~\cite{Kar01} for $m_q/T=1$, {\it i.e.}, 
$g=\sqrt 6$.
Generically the pole-pole term of the meson spectral function in 
HTL-approximation describes three physical processes: {\bf (1)} the 
annihilation of collective quarks, 
{\bf (2)} the annihilation of two plasminos,
and {\bf (3)} the transition from upper to lower branch.
The transition process starts at
zero energy and continues until the maximum difference $\omega =0.47\> m_q$
between the two branches at $k=1.18\> m_q$. At this point a Van Hove
singularity is encountered due to 
a diverging density of states for the third process as noted above.
The plasmino annihilation starts at $\omega = 1.86\> m_q$ with another Van 
Hove singularity corresponding to the minimum of the plasmino branch at 
$k=0.41\> m_q$, where again the density of states diverges corresponding 
to second process. 
This contribution falls off 
rapidly due to the exponentially suppressed spectral strength of the plasmino 
mode for large energies, where only the first process, quark-antiquark 
annihilation starting at $\omega =2m_q$, contributes. For large energies this
dominates and approaches the free results (crosses) for $\omega >> m_q$.
The pole-cut and cut-cut contributions, which involve 
external gluons as can be seen by cutting the HTL quark 
self energy in figure~\ref{fig:hself}, lead
to a smooth contribution to the spectral function. The 
pole-pole and pole-cut contributions in both channels are of similar
magnitude,
while the cut-cut contribution at small $\omega$ vanishes in the pseudoscalar 
channel but diverges in the vector channel. This has its origin in the 
structure of HTL quark-meson vertex~\cite{Bra90,Fre90}, which is 
required for the 
vector channel containing a collinear singularity, 
whereas bare vertices are sufficient for the (pseudo)scalar channel.
In a very recent calculation~\cite{nla} using next to leading order
HTL approximation, the pseudoscalar spectral function is found to remain
almost same as the leading order results.

\subsection{Dilepton production rate in HTL-approximation}
The static dilepton production rate (${\vec p}=0$) \cite{Bra90}
is related to the vector meson channel as
\beq
\sigma_V(\omega)=\frac{1}{\pi}{\rm{Im}}\ \chi_V(\omega)
=\frac{18\pi^2N_C}{5\alpha^2}
\left( e^{\beta \omega}-1\right ) \omega^2 \frac{dR}{d^4xd\omega d^3p}
({\vec p}=0) . \label{eq36}
\eeq

\begin{figure}[htbp]
\centerline{{\epsfxsize=6cm {\epsfbox{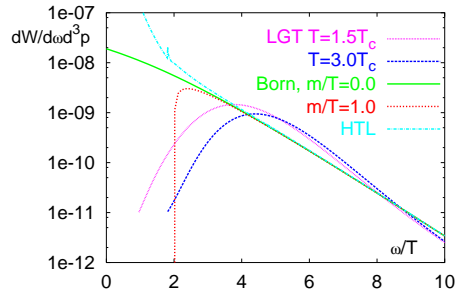}}}}
\caption{ The dilepton production rate
in HTL-approximation for $m_q/T=2$ is compared with Lattice data.}
\label{fig:dilep}
\end{figure}

In figure~\ref{fig:dilep} we show the dilepton rates calculated
from (\ref{eq36}) in HTL and also in Born approximations, which are
compared with
the first dilepton rate obtained in lattice~\cite{latdil}. 
The comparison of lattice rate with HTL and Born rate shows that for 
all energies 
$\omega/T\geq 4$ the difference is very small. For energies $\omega/T\leq 3$, 
the lattice dilepton rate drops rapidly and reflects sharp cut-off
found in reconstructed spectral function~\cite{latdil} and differs from that
of HTL rate.

\subsection{Meson correlation functions in HTL-approximation}
\begin{figure}[htbp]
\centerline{{\epsfxsize=6cm {\epsfbox{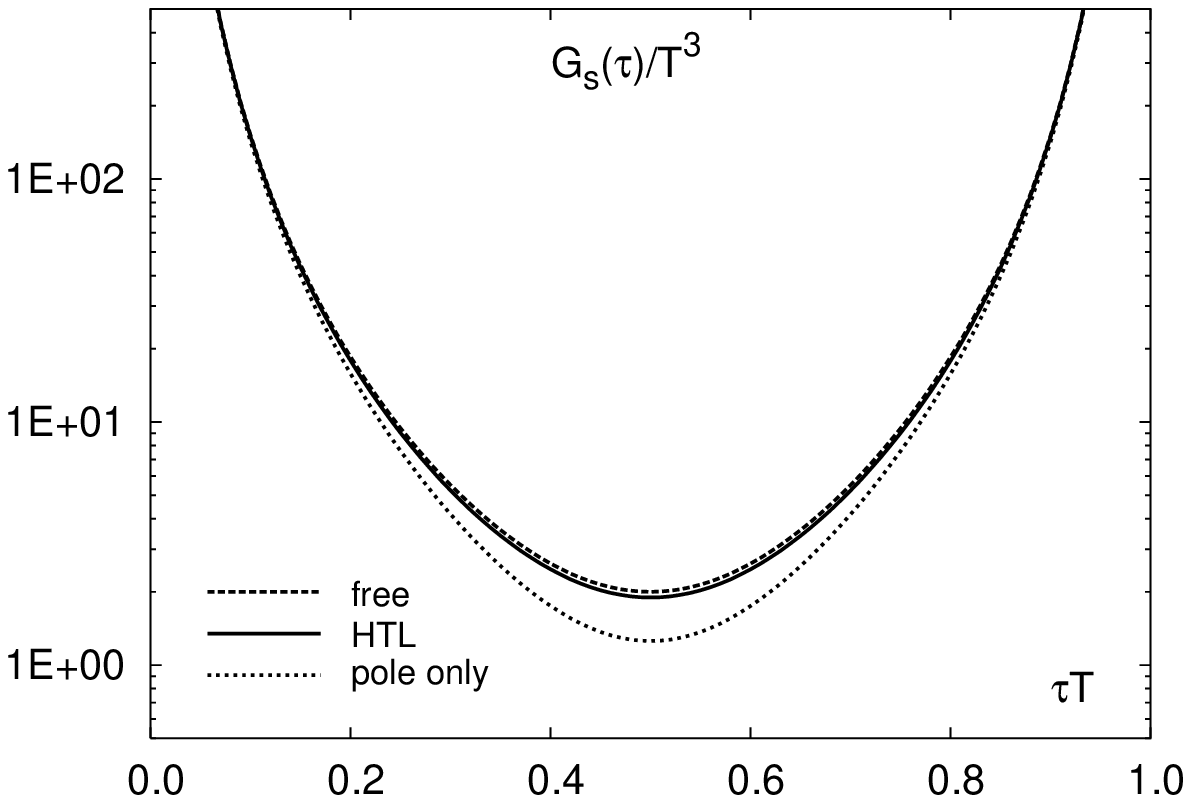}}}
{\epsfxsize=6cm {\epsfbox{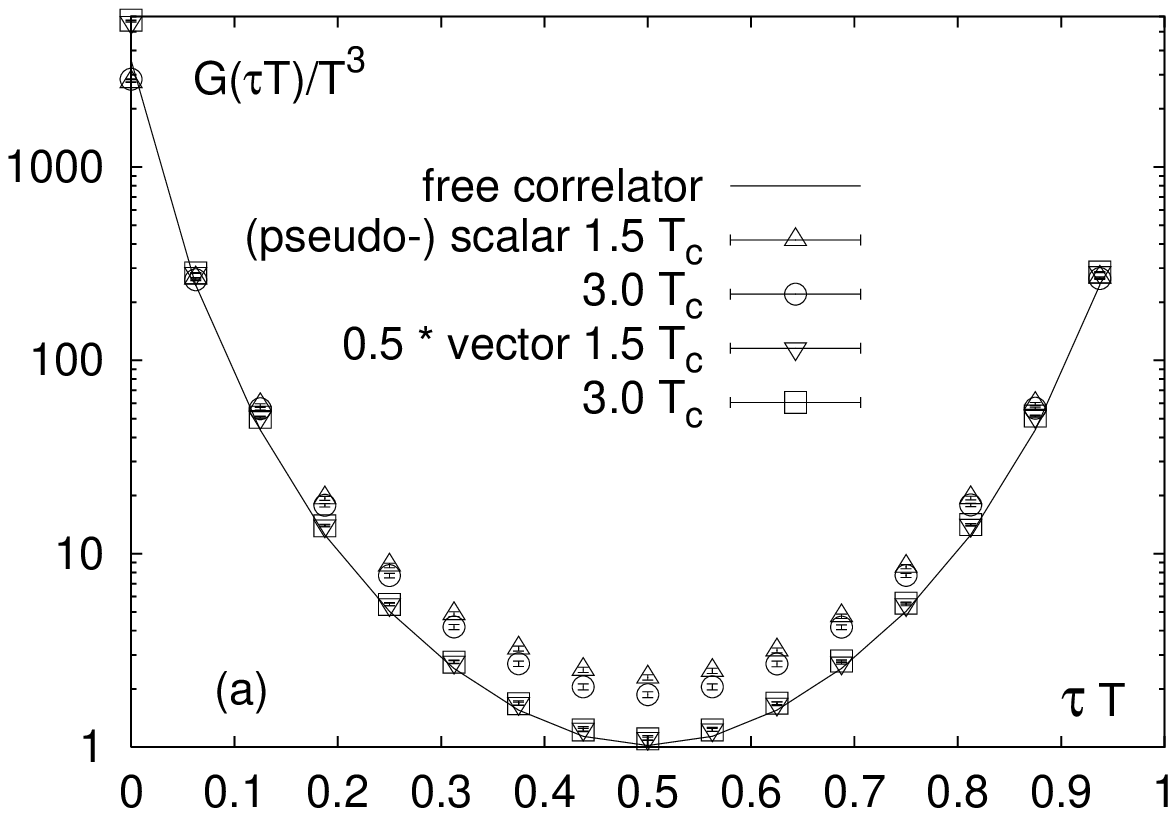}}}}
\caption{ The pseudoscalar 
meson correlation functions in HTL-approximation (left) for $m_q/T=1$ and
compared with lattice results.}
\label{fig:pscor}
\end{figure}
The temporal correlators for pseudoscalar and vector mesons can be
obtained from their spectral functions according to (\ref{eq35})
in order to take into account of medium effects from the quark propagator. 
At larger temperatures it seems that the pseudoscalar correlator
only slowly approaches the free correlation function and still differs  
from the lattice results~\cite{latdil}. In other words, the HTL medium effects
may not be sufficient to explain the deviation from free one. Details of
the vector meson correlation function can be found in Ref.~\cite{Kar01}.

\section{Susceptibilities}

Dynamical properties of a many particle system can be investigated by
employing an external probe, which disturbs the system only slightly
in its equilibrium state, and by measuring the response of the system to
this external perturbation. 
Usually one probes the dynamical behavior of the
spontaneous fluctuations in the equilibrium state. In general,
spontaneous fluctuations are related to correlation functions.
They reflect the symmetries of the system,
which provide important inputs for quantitative
calculations of complicated many-body systems.
Recently, screening and fluctuations of conserved quantities have been
proposed as a probe of the quark-gluon plasma (QGP) formation
in ultrarelativistic heavy-ion
collisions.

\subsection{Definition}
Let \( {\cal O}_{\alpha } \) be a Heisenberg operator. In a static and 
uniform
external field \( {\cal F}_{\alpha } \), the (induced)
expectation value of the operator 
\( {\cal O}_\alpha \left( 0,\overrightarrow{x}\right) \) 
is written as
\begin{eqnarray}
\phi _{\alpha }&\equiv& \left\langle {\cal O} _{\alpha }\left
( 0,\overrightarrow{x}\right) \right\rangle _{F}=\frac{{\rm Tr}\left
[ {\cal O} _{\alpha }\left( 0,\overrightarrow{x}\right) e^{-\beta \left
( {\cal H}+{\cal H}_{ex}\right) }\right] }{{\rm Tr}\left[ e^{-\beta
\left( {\cal H}+{\cal H}_{ex}\right) }
\right] }=\frac{1}{V}\int d^{3}x\, \left\langle {\cal O} _{\alpha }
\left( 0,\overrightarrow{x}\right) \right\rangle \: , \label{eq37}
\end{eqnarray}
where the translational invariance is assumed and
\({\cal H}_{ex} \) is given by
\({\cal H}_{ex}=-\sum _{\alpha }\int d^{3}x\, {\cal O} _{\alpha }\left( 0,
\overrightarrow{x}\right) {\cal F}_{\alpha }\) .
The (static) susceptibility \( \chi _{\alpha \beta } \) is defined as
                                                                                
\begin{eqnarray}
\chi _{\alpha \sigma }(T) & = & \left. \frac{\partial \phi _{\alpha }}
{\partial {\cal F}_{\sigma }}\right| _{{\cal F}=0}
  =  \beta \int d^{3}x\, \left\langle {\cal O} _{\alpha }\left
( 0,\overrightarrow{x}\right) {\cal O} _{\sigma }\left( 0,\overrightarrow{0}
\right) \right\rangle \: , \label{eq38}
\end{eqnarray}
assuming no broken symmetry \( \left\langle {\cal O} _{\alpha }
\left( 0,\overrightarrow{x}\right) \right\rangle =\left\langle
{\cal O} _{\sigma }
\left( 0,\overrightarrow{0}\right) \right\rangle =0 \).
$\langle {\cal O}_\alpha (0,{\vec x}){\cal O}_\sigma(0,{\vec 0})\rangle $
is the two point correlation function with operators evaluated
at equal times.
                                                                                
\subsection{{Quark number susceptibility:}}
\begin{figure}[htbp]
\centerline{{\epsfxsize=9cm {\epsfbox{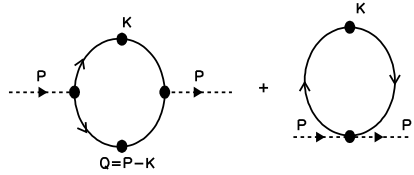}}}}
\vspace{-4.0cm}
\caption{The 1-loop vector meson self energy diagram.} 
\label{fig:qnsdia}
\end{figure}

The quark number susceptibility is the measure of the response of the
quark number density with infinitesimal changes in the quark chemical
potential $\mu_q +\delta \mu_q$ and can be written as
\begin{eqnarray}
\chi_q(T) = \left.\frac{\partial \rho_q}{\partial \mu_q}\right |_{\mu_q=0}
= \beta\int \ d^3x \ \langle j_0(0,{\vec x}) j_0(0,\vec 0) \rangle
= \beta\int \ d^3x \ S_{00}(0,{\vec x})
 \, \, , \label{eq39}
\end{eqnarray}
where $S_{00}(0,{\vec x})$ is the time-time component of 
 {{vector meson correlator}} 
$S_{\mu\nu}(t,{\vec x})= \, \langle
j_\mu(t,{\vec x})j_\nu(0,{\vec 0})\rangle$.
With the Fourier transform of
$S_{00} (0,\vec x)$ the (\ref{eq39}) becomes 
 \begin{eqnarray}
\chi _{q}\left( T\right) &=&
\beta \int _{-\infty }^{+\infty }\frac{d\omega }
{2\pi }\, S_{00}(\omega,0) \, \,  \, \,
{\rm{with}} \, \, \, 
S_{00}(\omega,p) =
\frac{-2}{1-e^{-\beta \omega }}\, \rm{Im}\,
\Pi _{00}\left( \omega , {\vec p}\right)
. \label{eq40}
\end{eqnarray}                                                                
Using the imaginary part of the 1-loop vector meson self-energy within the 
HTL-approximation given in figure~\ref{fig:qnsdia} the quark number 
susceptibility~\cite{purnqns} is computed through (\ref{eq40}).
It is found to contain~\cite{purnqns} only the quasiparticle (QP) 
contributions following from the QP dispersion relation $D_\pm=0$,
given in (\ref{eq31}). Note that the transversality condition in
figure~\ref{fig:qnsdia} eliminates the Landau damping (LD) contributions
originating from the space like region.

\vspace{-2cm}

\begin{figure}[htbp]
\centerline{{\epsfxsize=5cm {\epsfbox{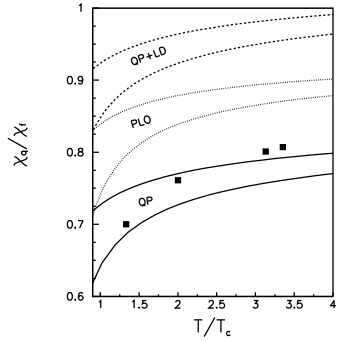}}}}
\vspace{-1cm}
\caption{The quark number susceptibility in
HTL approximation scaled with free susceptibility is plotted
as a function of $T/T_C$ with $T_C/\Lambda_{\bar{MS}}=0.69$.
In each band the lower curve corresponds to the choice of the renormalization
scale ${\bar {\mu}}=2\pi T$ and the upper one to ${\bar {\mu}}=2\pi T$. The
squares represent recent lattice data.} 
\label{fig:qns}
\end{figure}
 The quark number susceptibility is plotted in 
figure~\ref{fig:qns} and compared with the lattice results. 
The result (band between the solid line denoted by QP)
agrees well with recent lattice data~\cite{gavai1}. (Here we used
the 2-loop result for the running coupling constant $g(T)$. 
For the
renormalization scale $\bar \mu$ entering the running coupling constant,
we choose two different values leading to the bands.)
However,
it overcounts the leading order perturbative result (dotted band
denoted by PLO) given as
$\chi_q/\chi_f=(1-2\alpha_s/\pi+\cdots)$.
We note that this can be
cured by going to the calculation at 2-loop order. Recently, it has been shown
that the 2-loop order
calculation~\cite{andersen2} for the
thermodynamic potential reveals the correct inclusion of the leading order
effects but happens to be very close to the 1-loop result.
Now, employing only resummed propagators and bare vertices, the QNS
contains LD contributions coming from
the discontinuity of the 2-point HTL function in addition to the QP
contribution. The result is shown in figure~\ref{fig:qns} by
QP+LD. There are also calculations for quark number susceptibility
within the approximately self-consistent HTL approximations~\cite{rebhan}.

\subsection{Chiral susceptibility}
The chiral susceptibility measures the response of the chiral condensate to 
the infinitesimal change of the current quark mass $m+\partial m$. The  
static chiral susceptibility can be obtained as
\begin{eqnarray}
 \chi_c(T)= -\left.
\frac{\partial \langle {\bar q}q\rangle}{\partial m}\right |_{m=0}
&=&\beta \int {\rm d}^3x \; S(0, {\vec x}) = \beta \int_{-\infty}^{+\infty}
\frac{{\rm d}\omega}{2\pi} S(\omega,0) \, , \label{eq41} 
\end{eqnarray}
where $S(\omega,0)$ is the Fourier transformed  correlator of the scalar 
channel.

Note that, in contrast to the quark number susceptibility~\cite{purnqns} 
where the
zero temperature contribution vanishes due to the transversality properties 
of the vector channel, the static
chiral susceptibility contains a quadratic ultraviolet divergence coming 
from the zero temperature contribution. Using dimensional regularization 
the temperature
independent term disappears,
as there is no scale associated with it,
leading to $\chi_c^f(T)=-\frac{N_fN_c}{6}\> T^2$, where $N_f$ is the number 
of two light quark flavours and $N_C$ is the number of colour.
The next order contribution follows from the two-loop thermodynamic potential 
given in Ref.~\cite{kapus}. However, the second derivative of this expression 
with respect to $m$
diverges at $m=0$. Hence, the static chiral susceptibility cannot be 
calculated consistently
in usual perturbation theory beyond leading order,
but requires HTL resummation~\cite{purnchi}.

\vspace{-0.50cm}
\begin{figure}[htbp]
\centerline{{\epsfxsize=5cm {\epsfbox{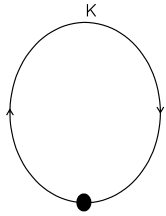}}}}
\vspace{-1cm}
\caption{The 1-loop scalar meson tadpole diagram.} 
\label{fig:chidia}
\end{figure}

\begin{figure}[htbp]
\centerline{{\epsfxsize=6cm {\epsfbox{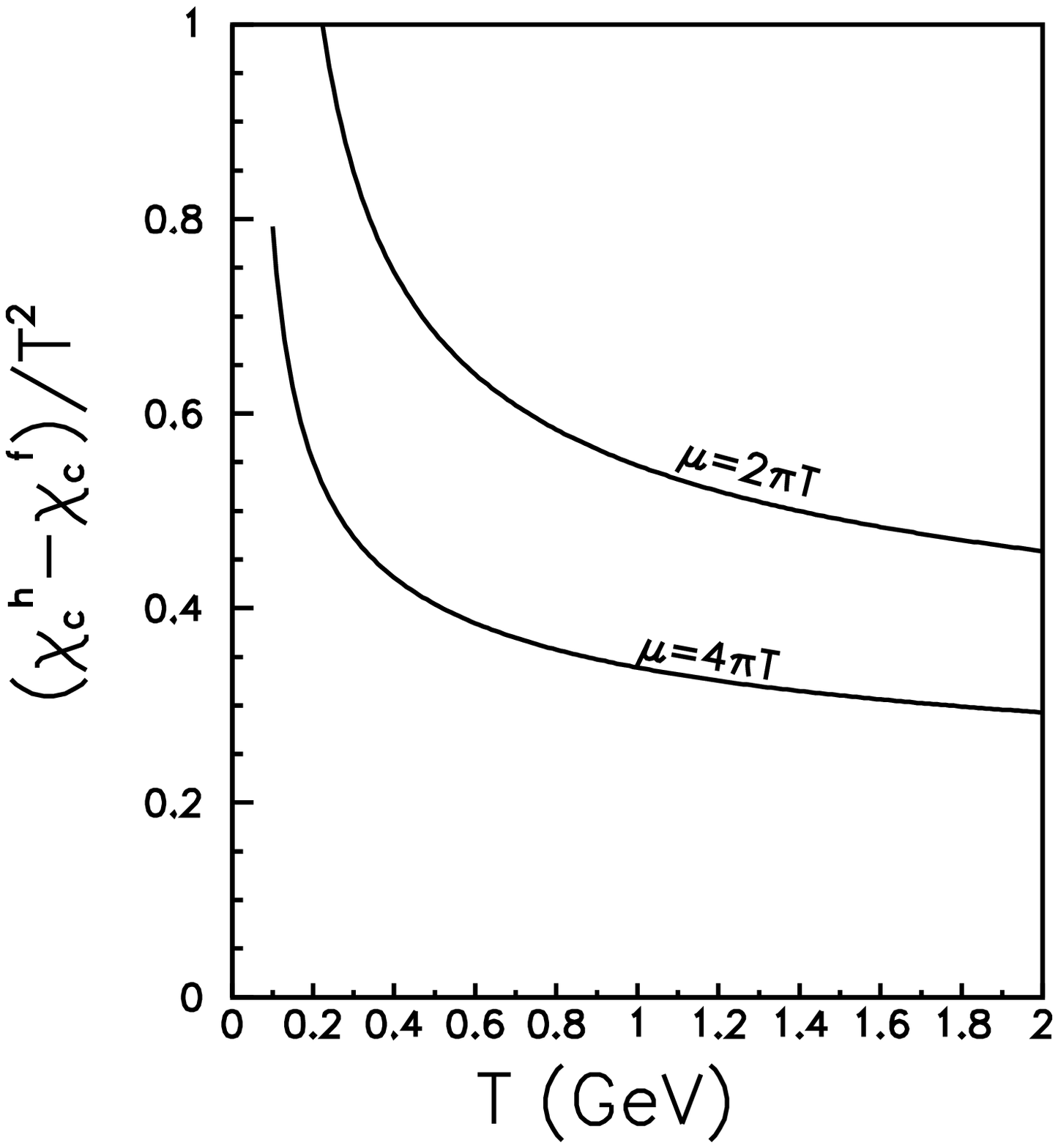}}}
{\epsfxsize=6cm {\epsfbox{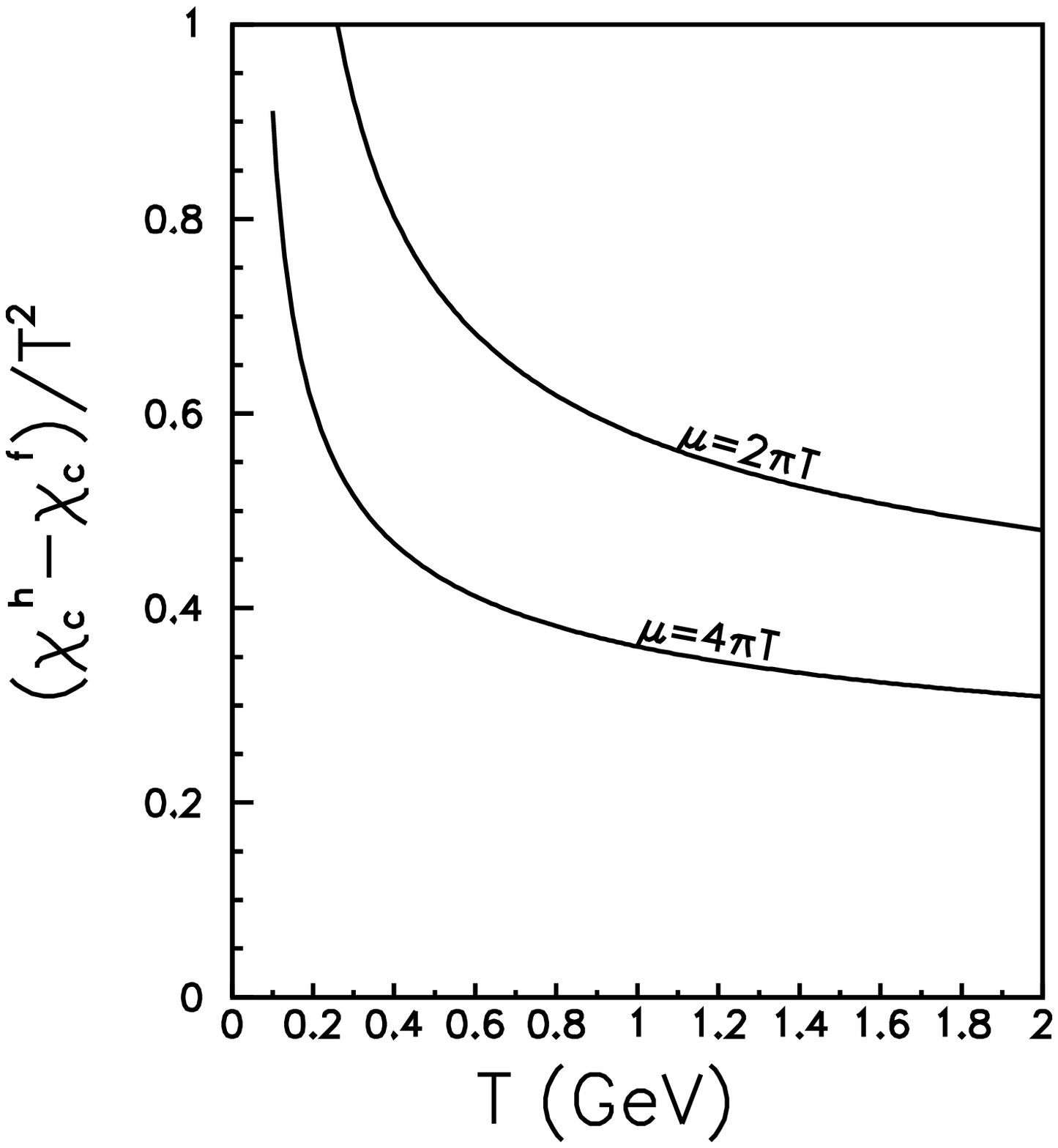}}}}
\vspace{-1cm}
\caption{ The HTL static susceptibility (free susceptibility subtracted)
as a function of temperature for $\Lambda_{\overline{\rm{MS}}}=300$ MeV,
and $N_f=2$ with the choices of the renormalization scale
$\mu=2\pi T$ and $4 \pi T$ for up to order $m_q^2$ (left) and
up to order $m_q^4$ (right).}
\label{fig:chi2}
\end{figure}

 Now, we compute the chiral condensate $\langle q\bar q\rangle$ from the
tadpole diagram in figure~\ref{fig:chidia}, where we use the effective 
HTL quark propagator,
yielding~\cite{purnchi}
\begin{eqnarray}
\frac{\chi_c^h(T)} {\chi_c^f(T)}&=& \left[ 1-\frac{6}{\pi^2}
\left (\log\frac{\mu}{4 \pi T}-\gamma -2\log 2 \right) \frac{m_q^2}{T^2} 
\right.
\left. +\frac{21}{4}\frac{\zeta(3)}{\pi^4}
\left(5+\frac{8}{7}\log 2 -\frac{\pi^2}{6} \right)
\frac{m_q^4}{T^4}
\right ] \, . \label{eq42}
\end{eqnarray}
Note that the $m_q^2$ term is the first finite $\alpha_s$ correction
to the free chiral susceptibility, since the two-loop correction in usual
perturbative QCD diverges for $m=0$. In figure~\ref{fig:chi2}, we show
the chiral susceptibility in HTL approximation. In lattice simulations a
peak around the critical temperature associated with chiral symmetry 
restoration has been observed. In contrast to lattice QCD the HTL
approximation, although taking into account the medium effects in plasma due to
the interactions, does not contain any physics related to the chiral
restoration, which is truly a nonperturbative effect. However, we
observe a strong increase of the magnitude of HTL chiral susceptibility
towards low temperature similar as in lattice simulations above $T_C$.

\vspace{-0cm}

\section{Dynamical Screening}
Screening of charges in a plasma is one of the most important collective
effects in plasma physics. In the classical limit in an isotropic
and homogeneous plasma the screening potential of a point-like
test charge $Q$ at rest can be derived from the
linearized Poisson equation, resulting in Debye screening. 
Earlier in most  calculations of the screening potential in the QGP, the
test charge was assumed to be at rest. However, quarks and gluons
coming from initial hard processes receive a transverse momentum which causes
them to propagate through the QGP. In addition, hydrodynamical
models predict a radial outward flow in the fireball. Hence,
it is of great interest to estimate the screening potential of a parton
moving~\cite{munshi} relatively to the QGP. Chu and Matsui \cite{Chu} have used
the Vlasov equation to investigate dynamic Debye screening for a heavy
quark-antiquark pair traversing a quark-gluon plasma. They found that the
screening potential becomes strongly anisotropic.
                                                                                
The screening potential of a moving charge $Q$ with velocity $v$ follows from
the linearized Vlasov and Poisson equations as
\begin{equation}
\phi({\vec r}, t; {\vec v}) = \frac{Q}{2\pi^2} \int d^3k
\frac{\exp{[-i{\vec k}\cdot ({\vec r}-{\vec v}t)]}}
{k^2 {\rm Re}[\epsilon_l (\omega={\vec k}\cdot {\vec v}, k)]}.
\label{e4}
\end{equation}
The dielectric function following from the semi-classical
Vlasov equation describing a collisionless plasma is related to
the high-temperature limit of the polarization tensor.
For example, the longitudinal dielectric function following from
the Vlasov equation is given by 
\begin{eqnarray}
\epsilon_l (\omega ,k) = 1-\frac{\Pi_{00}(\omega ,k)}{k^2}
= 1+\frac{m_D^2}{k^2} \left (1-\frac{\omega}{2k} \ln \frac {\omega +k}
{\omega -k}\right ),
\label{e3}
\end{eqnarray}
where the only non-classical inputs are
Fermi and Bose distributions instead of the Boltzmann distribution.
The gluon self-energies derived within the hard thermal loop
approximation have been shown to be gauge invariant
and the dielectric functions obtained from these are therefore
also gauge invariant.
%

In the general case, for parton velocities $v$
between 0 and 1, we have to
solve (\ref{e4}) together with (\ref{e3}) numerically. Since the potential
is not isotropic anymore due to the velocity vector ${\vec v}$,
we will restrict ourselves only to two cases, ${\vec r}$ parallel to ${\vec v}$
and ${\vec r}$ perpendicular to ${\vec v}$, i.e., for illustration
we consider the screening potential only in the direction
of the moving parton. The details of the perpendicular case can be
found in Ref.~\cite{munshi}.
                                                                                
\vspace{-0cm}
\begin{figure}[htbp]
\centerline{{\epsfxsize=5cm {\epsfbox{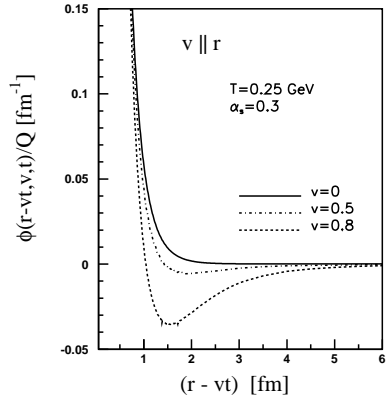}}}}
\vspace{-0cm}
\caption{
Screening potential parallel to the velocity of the moving parton
in a QGP.}
\label{fig:pot}
\end{figure}

In figure~\ref{fig:pot} the screening potential
$\phi /Q$ in ${\vec v}$-direction is shown as a function of $r'=r-vt$,
where $r=|{\vec r}|$, between 0 and 6 fm for various velocities.
For illustration we have chosen a strong
fine structure constant $\alpha_s =g^2/(4\pi)=0.3$,
a temperature $T=0.25$ GeV, and the number of quark flavors $N_f=2$.
The shifted potentials~\cite{purnchi} depend only on $v$ and
not on $t$ as it should be the case in a homogeneous and isotropic plasma.
For $r'<1$ fm one observes that the fall-off of the potential
is stronger than for a parton at rest.
The reason for this behavior is the fact that there is a stronger
screening in the direction
of the moving parton due to an enhancement of the particle density
in the rest frame of the moving parton.
                                                                                
In addition, a minimum in the screening potential at $r'>1$ fm shows up
due to the loss of spherical symmetry of the Debye screening cloud. For
example, for $v=0.8$ this minimum is at about 1.5 fm with a depth
of about 8 MeV.
 A minimum in the screening potential is also known from non-relativistic,
complex
plasmas, where an attractive potential even between equal charges
can be found if the finite extension of the charges is considered
\cite{Tsytovich}.
A similar screening potential was found for a color charge at rest
in Ref.\cite{Gale}, where a polarization tensor beyond the high-temperature
limit was used.
However, this approach has its limitation as a gauge dependent and incomplete
(within the order
of the coupling constant) approximation for the polarization tensor 
was used. 
Obviously, a minimum in the interparticle potential in a relativistic
or non-relativistic plasma is a general feature if one goes beyond the Debye-H\"uckel
approximation by either taking quantum effects, finite velocities, or finite sizes of the particles
into account.

The modification of the confinement potential below the critical temperature
into a Yukawa potential above the critical temperature
might have important consequences for the discovery
of the QGP in relativistic heavy-ion collisions. Bound states of
heavy quarks, in particular the $J/\psi$ meson, which are produced in the
initial hard scattering processes of the collision, will be dissociated in
the QGP due to screening of the quark potential and break-up by energetic
gluons. 
On the other hand, the formation of colored bound states, 
{\textit e.g.}, $qq$, $\bar q\bar q$, $gg$, of partons
at rest has also been claimed~\cite{Shuryak} above the critical
temperature (2$T_c$ - 3$T_c$) by analyzing lattice data,
indicating that the plasma behaves as a strongly coupled quark-gluon plasma.
The dynamical screening will have important consequences on the $J/\Psi$ 
dissociation and other binary states formed in the QGP just above $T_C$.

\section{Summary}
In this lecture the thermal field theory within the imaginary time formalism
has been introduced and some of the problems in computing
physical quantities in bare perturbation theory are pointed out. In order to
overcome these problems, the prescription of HTL resummation technique is
discussed and based on it an improved perturbation theory (HTLpt) is also 
briefly outlined. Using this improved HTLpt various physical
quantities, hadronic spectral and correlation functions, quark number and
chiral susceptibilities, and the dynamical screening potentials have been
computed and discussed at length. 

\acknowledgments
Most of the works were done in collaboration with
P. Chakraborty, F. Karsch and M. H. Thoma.


\begin{thebibliography}{99}
\bibitem{qm99}Proceeding of the {\sf 14th International Conference on
Ultra-Relativistic Nucleus-Nucleus Collisions-$``$ Quark Matter'99"},
Nucl. Phys A{\bf 661} (2000).
\bibitem{qm04}Proceeding of the {\sf 17th International Conference on
Ultra-Relativistic Nucleus-Nucleus Collisions-$``$ Quark Matter'04}",
J. Phys. G{\bf 30}(2004).
\bibitem{karsch0} F. Karsch, Lattice'99, Proceedings of the XVII{\it th}
International Symposium on Lattice Field Theory,
Nucl. Phys. B (Proc. Suppl.) {\bf 83-84}, 14$_c$ (2000).
\bibitem{kapus} J. I. Kapusta, {\it Finite-Temperature Field Theory} (Cambridge
University Press, Cambridge, 1989).
\bibitem{lebel} M. Le Bellac, {\it Thermal Field Theory} 
(Cambridge University Press, Cambridge, 1996).
\bibitem{Das} A.Das, {\it Finite Temperature Field Theory} (World Scientific,
Singapore, 1997) p. 102.
\bibitem{markus} M.H. Thoma, in {\it Quark-Gluon Plasma 2}, ed.: R.C. Hwa
(World Scientific, Singapore, 1995) p.51; M. H. Thoma, {\it New Developments
and Application of Thermal Field Theory}, $\langle$ hep-ph/0010164 $\rangle$.
\bibitem{Kaj85} K. Kajantie and J. I. Kapusta, Ann. Phys. (N.Y.) {\bf 160}
(1985) 477.
\bibitem{Lop85} J. C. Parikh, P. J. Siemens, and J.A. Lopez, 
Pramana {\bf 32} (1989) 555.
\bibitem{Bra90a} E. Braaten and R.D. Pisarski, Nucl. Phys. B {\bf 337}
 (1990) 569.
\bibitem{Fre90} J. Frenkel and J. C. Taylor, Nucl. Phys. B {\bf 334}
(1990) 199.
\bibitem{braaten0} E. Braaten and R. D. Pisarski, Phys. Rev. D{\bf 45}
(1992) R1827.
\bibitem{andersen} J. O. Andersen, E. Braaten, and M. Strickland,
Phys. Rev. Lett. {\bf 83} (1999) 2139.
\bibitem{karsch} F. Karsch, A. Patk\'os, and P. Petreczky, Phys. Lett.
B{\bf 401} (1997).
\bibitem{andersen0} J. O. Andersen, E. Braaten, and M. Strickland,
Phys. Rev. D {\bf 61} (2000) 014017.
\bibitem{andersen1} J. O. Andersen, E. Braaten, and M. Strickland,
Phys. Rev. D {\bf 61} (2000) 074016.
\bibitem{andersen1a} J. O. Andersen  and M. Strickland,
Phys. Rev. D {\bf 66} (2002) 105001.
\bibitem{andersen2} J. O. Andersen, E. Braaten, E. Petitgirard and M.
Strickland, Phys. Rev. D{\bf 66} (2002) 085016; J. O. Andersen, 
E. Petitgirard and M.  Strickland, Phys. Rev. D{\bf 70} (2004) 045001.
\bibitem{Bra90} E. Braaten, R.D. Pisarski, and T.C. Yuan, Phys. Rev. Lett.
{\bf 64}, (1990) 2242.
\bibitem{Wel82a} H.A. Weldon, Phys. Rev. D {\bf 26} (1982) 2789.
\bibitem{Kar01} F. Karsch, M. G. Mustafa, and M.H. Thoma, 
Phys. Lett. B {\bf 497} (2001) 249.
\bibitem{Flo94}W. Florkowski and B.L. Friman, Z. Phys. A {\bf 347}
 (1994) 271.
\bibitem{nla}W. M. Alberico, A. Beraudo and A. Molinari, 
Nucl. Phys. A {\bf 750} (2005) 359.
\bibitem{latdil}F. Karsch et al, Phys. Lett. B {\bf 530} (2002) 147.
\bibitem{purnqns} P. Chakraborty, M. G. Mustafa, Euro. Phys. Jour. C
{\bf 23} (2002) 591; Phys. Rev. D {\bf 68} (2003) 085012.
\bibitem{gavai1} R. V. Gavai and S. Gupta, Phys. Rev. D {\bf 65} (2002)
094515; R. V. Gavai, S. Gupta, and P.
Phys. Rev. D {\bf 65} (2002) 054506.
\bibitem{rebhan}J.-P. Blaizot, E. Iancu and A. Rebhan, Phys. Lett. B {\bf 523}
(2001) 143; Euro. Phys. Jour. C {\bf 27} (2003) 433.
\bibitem{purnchi} P. Chakraborty, M. G. Mustafa and M. H. Thoma, 
Phys. Rev. D {\bf 67} (2003) 114004.
\bibitem{munshi} M. G. Mustafa, M. H. Thoma and P. Chakraborty, Phys. Rev. C
{\bf 71} (2005) 017901.
\bibitem{Chu}M. C. Chu and T. Matsui, Phys. Rev. D {\bf 39} (1989) 1892.
\bibitem{Tsytovich} V. N. Tsytovich, JETP Lett. {\bf 78} (2003) 1283.
\bibitem{Gale} C. Gale and J. Kapusta, Phys. Lett. B {\bf 198} (1987) 89.
\bibitem{Shuryak} E V Shuryak and I Zahed, Phys. Rev D{\bf 70} (2004) 054507.
\end{thebibliography}
\end{document}